\documentclass{sig-alternate-10pt}
\usepackage[utf8]{inputenc}
\usepackage[T1]{fontenc}
\usepackage{tcolorbox}
\usepackage{graphicx}
\graphicspath{{figures/}}
\usepackage[hyphens]{url}
\usepackage{hyperref}
\usepackage[absolute,showboxes]{textpos}
\usepackage{booktabs}
\usepackage{fancyvrb}
\usepackage{framed}
\usepackage{lastpage}
\usepackage[multiple]{footmisc}
\usepackage[obeyDraft]{todonotes}
\presetkeys
	{todonotes}
	{inline}{}
\newcommand*{\refname}{}
\def\sharedaffiliation{
\end{tabular}\\
\begin{tabular}{c}}
\numberofauthors{3} 
\author{
	\alignauthor
	Timm B\"ottger
	\email{timm.boettger@qmul.ac.uk}
	\alignauthor
	Felix Cuadrado
	\email{felix.cuadrado@qmul.ac.uk}
	\alignauthor
	Gareth Tyson
	\email{gareth.tyson@qmul.ac.uk}
	\and
	\alignauthor
	Ignacio Castro
	\email{i.castro@qmul.ac.uk}
	\alignauthor
	Steve Uhlig
	\email{steve.uhlig@qmul.ac.uk}
	\sharedaffiliation
	\affaddr{Queen Mary University of London}
}
\title{Open Connect Everywhere: A Glimpse at the Internet Ecosystem through the Lens of the Netflix CDN}
\begin{document}
\setlength{\TPHorizModule}{\paperwidth}
\setlength{\TPVertModule}{\paperheight}
\TPMargin{5pt}
\begin{textblock}{0.8}(0.1,0.02)    
     \noindent
     \footnotesize
     If you cite this paper, please use the CCR reference:
     B\"ottger, T., Cuadrado, F., Tyson, G., Castro, I., Uhlig, S.. (2018). "Open Connect Everywhere: A Glimpse at the Internet Ecosystem Through the Lens of the Netflix CDN". In ACM SIGCOMM Computer Communications Review (CCR), 48(1).
\end{textblock}
\maketitle
\begin{abstract}
The importance of IXPs to interconnect different networks and exchange traffic locally has been well studied over the last few years. However, far less is known about the role IXPs play as a platform to enable large-scale content delivery and to reach a world-wide customer base. In this paper, we study the infrastructure deployment of a content hypergiant, Netflix, and show that the combined worldwide IXP substrate is the major corner stone of its Content Delivery Network.
To meet its worldwide demand for high-quality video delivery, Netflix has built a dedicated CDN. 
Its scale allows us to study a major part of the Internet ecosystem, by observing how Netflix takes advantage of the combined capabilities of IXPs and ISPs present in different regions.
We find wide disparities in the regional Netflix deployment and traffic levels at IXPs and ISPs across various local ecosystems. This highlights the complexity of large-scale content delivery as well as differences in the capabilities of IXPs in specific regions.
On a global scale we find that the footprint provided by IXPs allows Netflix to deliver most of its traffic directly from them.
This highlights the additional role that IXPs play in the Internet ecosystem, not just in terms of interconnection, but also allowing players such as Netflix to deliver significant amounts of traffic.
\end{abstract}
\section{Introduction}
\label{sec:introduction}
Originally designed as a research network, the Internet has evolved into a massive-scale platform for multimedia delivery. This transformation has been possible thanks to many underlying technical evolutions and innovations, stretching the Internet way beyond its original design. In this paper, we focus on two such shifts that are dramatically impacting the way the Internet operates today. First, a topological flattening has been observed~\cite{gill2008flattening}, driven partly by the expansion of Internet Exchange Points (IXPs). These IXPs commoditise the interconnection of networks~\cite{labovitz2010internet}, and significantly lower the cost of network operations. Previous studies uncovered a rich and varied network ecosystem inside an IXP, so large that it fundamentally questions our current knowledge of the AS-level topology~\cite{Ager2012IXP}. Second, consumption of online content, especially video material, has steadily grown, sparking the deployment of content delivery infrastructures deep inside the network, e.g., ISP caches, on a global scale.
When combining the above two observations, we begin to see a greater emphasis on traffic being generated and exchanged locally, rather than following the traditional hierarchy. This process, led by so-called hypergiants~\cite{labovitz2010internet} (e.g., Google, Facebook), has radically altered the location of network ``hot spots'', reducing the importance of the traditional tier-1 networks and re-asserting the edge as the principal playground for innovation. Although previous studies have shown that individual IXPs are important for today's network interconnection landscape~\cite{Ager2012IXP, labovitz2010internet}, there yet is no thorough analysis of the role the IXP ecosystem plays to support major content delivery players. 
One of these major players or hypergiants is Netflix. Since 2012, Netflix has been deploying its own content delivery infrastructure, named Open Connect. It relies on server locations near to the edge, strategically located close to its user base. In contrast to other hypergiants (e.g. Google, Facebook), Netflix operates neither a backbone network nor datacenters~\cite{netflix_cloud_migration, netflix_peering_locations}. Instead Netflix pre-loads content on its servers during off-peak times to reduce the need for transit traffic~\cite{netflix_oc_fill}.
In this paper, we have performed the first large-scale measurement study of the Open Connect infrastructure. Using a range of techniques, we have discovered servers present at locations around the world and quantified the traffic generated by each of the appliances. Using location information provided in the server names, we study the regional footprints of the deployed infrastructure and expose a variety of regional Internet ecosystems. Our results not only reveal the dependence that Netflix has on these regional ecosystems, but also highlight the combined ability of the many IXPs world-wide to deliver huge amounts of traffic on a local scale. They bypass the traditional tier-1 and transit networks, thus underpinning the fact that hypergiants like Netflix contribute to the flattening of the Internet.
To summarise, in this paper we make the following contributions:
\begin{enumerate}
	\item We describe the infrastructure deployment of a content hypergiant (Netflix), which delivers large amounts of traffic from over 500 locations world-wide.	
	\item We provide evidence for the vastly understated ability of the many IXPs world-wide to deliver large amounts of traffic on a global scale. The world-wide footprint of IXPs enables Netflix to operate a global content delivery system, with very limited transit traffic, and without operating a backbone or owning datacenters. This complements previous observations regarding the importance of the IXP ecosystem for network interconnections.
	\item We expose the significant diversity of local ecosystems that collectively make up the Internet by highlighting the regional differences in Netflix's deployment. 
\end{enumerate}
We start the paper by providing background on Netflix (Section~\ref{sec:background}) and explaining our measurement methodology (Section~\ref{sec:methodology}).
Afterwards, we leverage the footprint of Netflix's infrastructure to expose the role of IXPs and a diversity of local ecosystems (Sections~\ref{sec:deployment} and \ref{sec:traffic}). We then put this work into perspective to the corpus of existing work (Section~\ref{sec:related}) and discuss how and why the derived view influences the overall view of the Internet (Section~\ref{sec:discussion}).
The paper ends with the summary (Section~\ref{sec:summary}).
 \section{Background}
\label{sec:background}
In this section, we provide the necessary background on Netflix's infrastructure and the data we collected.
We first highlight the characteristics that make Netflix a worthy object of study (Section~\ref{sec:why_study_Netflix}) 
and assess afterwards which parts of the Internet ecosystem to focus on (Section~\ref{sec:the_netflix_lens}).
\subsection{Why study Netflix?}
\label{sec:why_study_Netflix}
The main reason for looking into Netflix is its reach and scale. Netflix is a global provider of Video-on-demand services, available in all countries except for China, Crimea, North Korea and Syria~\cite{netflix_countries}. Netflix also is one of the largest content players on the market, being responsible for a significant amount of Internet traffic. Its global reach and size alone make it a valuable object of study.
Netflix previously relied on third-party CDNs for delivering content~\cite{adhikari2015measurement}.
However, it has recently abandoned this approach and developed its own CDN called Open Connect~\cite{netflix_openconnect_blog_map}.
CDN servers, called Open Connect appliances, are installed at IXPs or inside the networks of ISPs.
Having all content delivery infrastructure under its control, Netflix provides an opportunity to get a global view on a slice of the Internet, by studying a single but world-wide infrastructure. In particular, the risk in missing parts of the infrastructure is limited, compared to an infrastructure that is outsourced and hidden behind third-parties. In addition, the server deployment is largely driven by the demand for Netflix content. Indeed, Netflix does not deliver content for other players through its infrastructure, which for third-party CDNs would likely be the case and would make the results of the study more difficult to interpret.
Netflix promotes the deployment of Open Connect appliances as an opportunity for ISPs to localise Netflix traffic and to reduce their dependency on transit providers~\cite{netflix_openconnect}.
As an alternative, Netflix offers to serve content through direct peering links.
Hence, there is little value for Netflix to rely on third-party transit providers or even CDNs on a grand scale\footnote{For management purposes and content updates, some sort of global connectivity through transit providers is still obviously required. Quoting Netflix~\cite{netflix_openconnect_blog_map}: "Globally, close to 90\% of our traffic is delivered via direct connections between Open Connect and the residential Internet Service Providers (ISPs) our members use to access the internet."}. We thus expect the CDN infrastructure of Netflix to be preferentially close to the edge, i.e., close to Internet service providers and end users. 
We note that Netflix's deployment is probably biased towards ISPs with large number of Netflix subscribers, as there is a certain setup overhead for each installed server. Consequently, the sample of the Internet obtained through Netflix will exhibit the same bias.
However, this is actually advantageous for our study, as it tends to move the focus on networks with sufficiently large customer bases, therefore better sampling the eyeball part of the Internet.
\subsection{The Netflix Lens}
\label{sec:the_netflix_lens}
\begin{figure}
    \centering
    \includegraphics[width=\columnwidth]{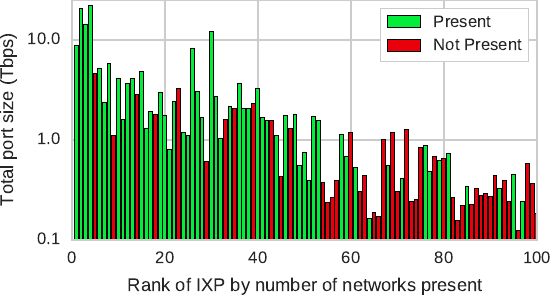}
    \caption{IXPs at which Netflix is present according to PeeringDB. The plot is limited to the 100 largest IXPs (in number of networks) for the sake of readability.}
    \label{fig:ixp-capacity-global}
\end{figure}

To better quantify the reach of Open Connect, we first inspect at which IXPs Netflix is present.
We rely on IXP data from PeeringDB~\cite{peeringdb} 
and determine the relative importance of an IXP by counting the number of networks that are connected to it.
We then rank IXPs by decreasing number of networks present.
Every network is only counted once, even if it has multiple routers connected at a single IXP.
Furthermore, IXPs not having a single network present according to PeeringDB are discarded.
Figure~\ref{fig:ixp-capacity-global} depicts which IXPs Netflix is present at.
On the x-axis, IXPs are ranked by decreasing numbers of networks present. On the y-axis, the height of each bar indicates the capacity of each IXP in Tbps, calculated as the sum over the reported capacity of all individual peering ports.
We observe two things from Figure~\ref{fig:ixp-capacity-global}.
First, Netflix is present at the four largest IXPs in number of networks connected worldwide (AMS-IX, IX.br São Paulo, DE-CIX and LINX). These four largest IXPs each have more than 650 connected networks. In contrast, the next largest has barely 300. Netflix is present at nine of the top ten largest IXPs, and 38 of the top 50. We also observe that outside the top 50 IXPs, Netflix is present at only 16 of the remaining IXPs in the top 100.
Second, we observe that those IXPs with the largest number of networks, are not necessarily those with the largest capacity, arguably adding dimensions to be considered by Netflix and making their IXP selection process more complicated. Nevertheless, capacity-wise, Netflix is present at all of the top five IXPs, nine of the top ten and 37 of the top 50. Of the remaining IXPs in the 50--100 position, Netflix is present only at 17. This indicates a clear bias from Netflix to choose its presence preferentially among larger IXPs in number of members as well as in capacity, and less among smaller IXPs. As Netflix preferentially chooses those large IXPs, we can expect to sample the important IXPs in a given region by studying Open Connect.
PeeringDB, however, does not allow us to assess the penetration of appliances into ISPs. We will thus revisit this question in Section~\ref{sec:dataset_overview}.
\section{Methodology}
\label{sec:methodology}
In this section we describe the methodology we use to discover servers deployed by Netflix. We briefly describe the relevant implementation details of the Open Connect infrastructure (Section~\ref{sec:delivery_infrastructure}), before describing the actual collection process in detail (Section~\ref{sec:data_collection}). Afterwards we validate the obtained data (Section~\ref{sec:data_validation}) and provide a first overview of the results (Section~\ref{sec:dataset_overview}).
\subsection{Open Connect Infrastructure}
\label{sec:delivery_infrastructure}
Netflix uses Amazon Web Services (AWS) for many of its computing tasks. Such computing tasks for example are serving of the website, the main application logic and the recommendation system, but also tasks related to video pre-processing and transcoding. The actual video content however is exclusively delivered through Netflix's own CDN Open Connect~\cite{netflix_cloud_migration}. It is only this delivery infrastructure that we examine in this study.
To better understand how individual video clients are assigned to content servers, we ran a measurement campaign using HTTP proxies from a multitude of vantage points.  We used the browser plugin Hola for this, which gave us vantage through 753 different IPs in 94 ASes. When a client requests a video file, the main application logic directly instructs the client which content servers to use. It (typically) hands out three domain names. The client then directly requests the video content from these servers via HTTPs.
The server names are very specific. They include information on the physical cache location and a cache number.
This detailed naming structure makes it unlikely that names resolve to more than one IP address.
This is consistent with what Netflix publishes on the naming convention of servers~\cite{netflix_openconnect_portal_naming}.
Nevertheless, we used Planetlab to confirm that each name only resolves to a single and always the same IP address, independent of the client's location.
These findings, although more detailed, are in line with what Netflix publishes on how client redirection works~\cite{netflix_openconnect_deploymentguide}. Examples of server names used by Netflix are shown in Figure~\ref{fig:netflix_server_names}.
\begin{figure}[htb!]
	\fontsize{9pt}{10pt}
	\begin{Verbatim}[frame=single]
 ipv4_1-lagg0-c020.1.lhr001.ix.nflxvideo.net
 ipv6_1-lagg0-c002.1.lhr005.bt.isp.nflxvideo.net
	\end{Verbatim}
	\vspace{-5mm}
	\caption{Examples of Netflix server names.}
	\label{fig:netflix_server_names}
\end{figure}
We conjecture that the meaning of the individual components of a name are as described in Figure~\ref{fig:netflix_server_name_components}. We will revisit the correctness of these assumptions later in this section.
\begin{figure}[htb!]
	\begin{framed}
	\begin{description}
		\itemsep0em 
		\item[\textbf\texttt{ipv4} / \textbf\texttt{ipv6}:] IP protocol version.
		\item[\textbf\texttt{lagg0}:] Type of network card. We also found other NICs (i.e., cxgbe0, ixl0, mlx5en0, mce0).
		\item[\textbf\texttt{c020}:] Server counter for a given location.
		\item[\textbf\texttt{lhr001}:] IATA airport code of a location with counter.
		\item[\textbf\texttt{bt.isp} / \textbf\texttt{ix}:] Network (type) identifier; server operated inside ISP British Telecom or at an IXP
	\end{description}
	\end{framed}
	\vspace{-5mm}
	\caption{Components of a Netflix server name.}
	\label{fig:netflix_server_name_components}
\end{figure}
For the remainder of this paper, we will use the IATA airport code to infer the physical location of a server and the network identifier to distinguish between ISP and IXP servers\footnote{Netflix does not distinguish between public IXPs and private peering facilities, but qualifies both as IXPs via the 'ix' part of the server names. This is reasonable if both options are viewed as just a means for delivering traffic. For the remainder of this paper we will adopt this view as well.}.
Whenever we refer to the location of a server, we will use the airport code only without the counter, i.e., for three servers deployed at \textit{lax001}, \textit{lax002} and \textit{lax003}, the location will be \textit{lax} only, and the location \textit{lax} will have three servers deployed.
\subsection{Crawling DNS}
\label{sec:data_collection}
To unveil the Open Connect network, we use a DNS crawler which enumerates and tries to resolve all domain names matching the above scheme. If a domain name can be resolved to an IP address, we assume that we found a Netflix server.
Note that ignoring the structured nature of the names and simply iterating over all possible character sequences is practically infeasible and not desirable.\footnote{Assuming an alphabet of 26 characters plus ’.’, ’-’, ’ ’ as special characters and a prefix length of at least 30 characters (c.f. Fig.~\ref{fig:netflix_server_names}), enumerating all $29^{30}$ possible combinations in one year’s time would require roughly $2^{36}$ DNS queries per second.}
To narrow down the search space and limit the load on the DNS servers, the crawler is fed with lists of airport codes and ISP names, so that only DNS names for valid airports codes and ISPs are constructed. We further limit the number of probed DNS names, if no IP address is retrieved for a specific location and network operator.
We also rely on DNS server behaviour standardised in RFC 8020~\cite{draft-ietf-dnsop-nxdomain-cut} to prune empty DNS subtrees with a single query.
We used the following data sources to generate the input lists of airport codes and ISP names fed to the crawler:
\begin{description}
	\item[Wikipedia] We relied on Wikipedia to compile a list of IATA airport codes. While Wikipedia also has information on ISPs, extracting this information from Wikipedia is much more cumbersome, as it is spread across many pages and summary pages often are not updated frequently. We thus used additional sources to compile a list of ISPs.
	\item[Certificate Transparency] In the specific case of Netflix, we can leverage the Certificate Transparency (CT) project, to generate a list of relevant ISP names. The Google-driven project aims to increase Internet security by providing datastores of all issued SSL/TLS certificates, which are distributed amongst independent entities and cryptographically secured~\cite{certificate_transparency}. These datastores  allow individuals to verify certificate issuance. They can be used, for example, to detect rogue certificates issued without a genuine certificate request.
The peculiarity of Netflix to use subdomains for the airport code and network (type) identifier, requires their servers to use separate SSL/TLS certificates for each server location\footnote{A wildcard SSL/TLS certificate issued for *.nflxvideo.net will not be accepted as valid for the actual server domains~\cite{rfc2818}.}. These certificates are committed as individual log entries to CT. We can use these log entries to infer ISP names and airport codes used by Netflix.
In addition, Google, through the CT project, discovered a non-authorised pre-certificate for its domains issued by Symantec's Thawte CA~\cite{google_ct_symantec}. As a consequence, Google requested Symantec to log all issued certificates with CT.
As Netflix uses Symantec certificates for all its video delivery servers, we expect the CT logs to have complete coverage on the certificates used by Netflix's video delivery servers.
	\item[Peering DB] To cope with the unlikely event that an ISP is not discoverable by using certificate logs as outlined above, we extracted all network names from PeeringDB. We used these names and all subsets of them as possible inputs for our ISP list.
\end{description}
We furthermore ensured to have included at least all arguably important ISPs in all regions by manually checking websites, ISP rankings and comparison sites, and discussions in web forums. Nevertheless, we could not identify a single ISP which has Netflix servers deployed and did not show up in the CT logs, suggesting that they constitute a form of ground truth. 
We neglect IPv6 servers for this study, to avoid measurement artefacts introduced by dual-stacked servers. Errors could potentially be introduced in cases where an IPv4 and an IPv6 address are used by the same server. Given the prevalence of IPv4 over IPv6 connectivity, it is reasonable to assume, that the IPv6-enabled servers are a subset of the IPv4 ones. Consequently, every server with an IPv6 address also has an IPv4 address and neglecting IPv6 does not adversely affect our data collection. This assumption is in line with Netflix's policy on dual-stacking servers. Hence for the data collection, we restrict our attention on IPv4 addresses only. We do not query for domain names starting with \texttt{ipv6} and we look for \texttt{A} (not \texttt{AAAA}) records only.
 Unless explicitly stated otherwise, the data used in this paper was collected on May 15 2017. 
\subsection{Data Validation}
\label{sec:data_validation}
To complement our CT logs ground truth, we can use a map by Netflix of their Open Connect infrastructure, published in a blog entry~\cite{netflix_openconnect_blog_map} dating from March 2016. Our measurements are highly consistent with this map.
A comparison of the two makes it obvious that we in general observe the same global coverage and relative weight of individual regions.
However, our measurements were done more than a year after Netflix's data release and show significant additions and developments in certain regions.
Netflix's data only reveals qualitative information, while our measurements yield quantifiable results.
Furthermore, our measurements identify the ISP networks where Netflix has deployed servers.
All in all, we are confident that we observed a complete enough part of Netflix's video delivery infrastructure, allowing us to draw conclusions for those regions of the world, in which Netflix has a significant presence. For the following sections we will thus treat our data as a ground truth on Open Connect.
\subsection{Data Overview}\label{sec:dataset_overview}
\begin{figure}
    \centering
    \includegraphics[width=\columnwidth]{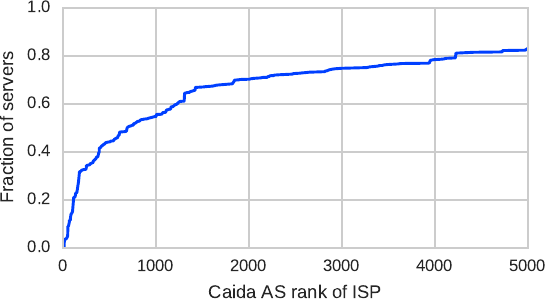}
    \caption{CAIDA AS rank of ISPs with Netflix servers deployed.}
    \label{fig:isp-caida-as-rank}
\end{figure}

\begin{figure*}
    \centering
    \includegraphics[width=\textwidth]{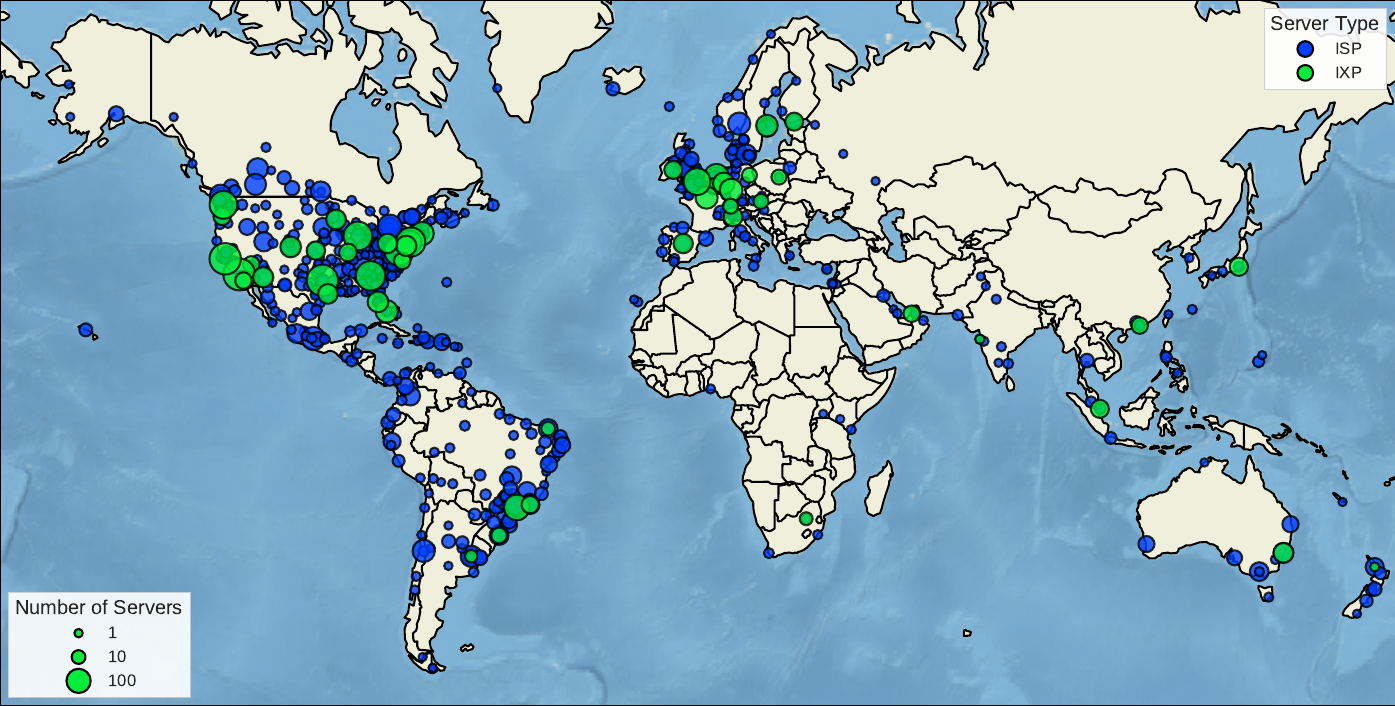}
    \caption{Netflix server deployment. Each marker denotes a location, the marker sizes are scaled by deployment size.}
    \label{fig:netflix-worldmap}
\end{figure*}

\begin{table}
    \centering
    \begin{tabular}{l|rr|r}
    \toprule
        ~ & ISP & IXP  & total \\ \midrule
        Servers & 4,152 & 4,340 & 8,492 \\
        Locations & 569 & 52 & 578 \\
        ASNs & 743 & 1 & 744 \\
        ISP names & 700 & - & - \\
    \bottomrule
    \end{tabular}
    \caption{Data Set Overview.}
    \label{tab:data-overview}
\end{table}
An overview of the gathered data set is shown in Table~\ref{tab:data-overview}.
In total we discovered 8,492 servers, of which 4,340 (51\%) are deployed within IXPs and 4,152 are deployed in ISPs. We observe servers at 569 different ISP and 52 different IXP locations, where a single location is a single airport code (see also Section~\ref{sec:delivery_infrastructure}). Our measurements reveal servers inside 700 different ISPs. While the IPs of all IXP servers are announced by the same AS, the IPs of the ISPs servers are announced by 743 ASs (which is more than the number of ISPs we observe). This happens because some ISPs use multiple AS numbers.
We can now also gauge the reach that Open Connect has amongst ISPs, which we could not do previously in Section~\ref{sec:the_netflix_lens}.
As an indicator for the importance of an ISP, we will use Caida's AS rank~\cite{caida_as_rank}. We used Team Cymru's IP to ASN mapping~\cite{teamcymru_ipasnmapping} to resolve IP addresses of the Open Connect servers to AS numbers. In those cases where we encountered multiple AS numbers for a single ISP, we used the AS number with the highest rank according to Caida. In Figure~\ref{fig:isp-caida-as-rank}, we show the CDF of the fraction of Netflix servers deployed across ISPs, ordered using the Caida's AS ranking. We observe that 50\% of the Netflix servers are deployed inside the top 500 ISPs according to Caida's AS rank. The other half of the servers are scattered across multiple thousands of ISPs. With the top 5000 ISPs, barely more than 80\% of Netflix servers are covered. Netflix ISPs servers are thus spread across a broad range of ISP networks, though preferentially amongst networks standing higher in Caida's AS ranking. 
Comparing the sheer number of ISP networks versus the relatively fewer IXPs where Netflix servers are deployed, we can already conclude that Netflix strategically chooses the IXPs where it is present, which are relatively few in numbers. This is in contrast to ISP deployments, where its servers are scattered across hundreds of ISPs. From this, we can expect very different granularities in Netflix IXP and ISP deployments, with fine-grained deployment in ISPs, while IXP deployments are likely to be more significant in terms of number of servers.
These different granularities also appear when looking at the geographical footprint of Open Connect. Figure~\ref{fig:netflix-worldmap} shows a plot of the server locations on a world map. Green dots indicate an IXP server location, blue dots indicate an ISP server location. The marker sizes are scaled by the number of servers at a location. Although Netflix offers its service globally, its servers are predominantly present in Western countries, their deployment mostly focuses on the Americas and Europe, and to a smaller extent on Australia.
\begin{figure}
    \centering
    \includegraphics[width=\columnwidth]{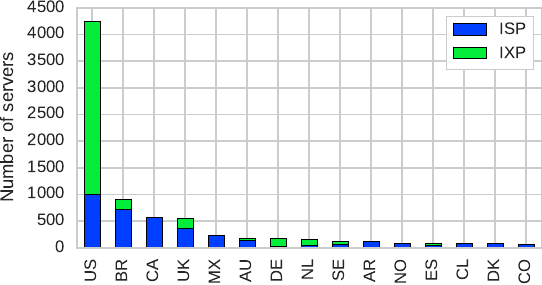}
    \caption{Countries with the largest Netflix deployments.}
    \label{fig:servers-per-country-top-15}
\end{figure}
Figure~\ref{fig:servers-per-country-top-15} depicts countries with the fifteen largest deployments of Open Connect servers. The largest deployment, by far, with 4,243 servers is in the US, followed by 901 servers in Brazil and 565 servers in the Canada. The United Kingdom and Mexico complete the top five countries\footnote{The complete dataset with server counts for all countries is available at \url{http://bit.ly/2rrn25S}.}.
 \section{The Local Ecosystems of the\\Internet}
\label{sec:deployment}
In this section, we describe the infrastructure deployment by Netflix in more detail. Our goal is to illustrate the diversity of the various local ecosystems that are part of the Internet and assess the role of IXPs in each ecosystem. We look at the largest deployments of Netflix servers in each continent, and expose different types of deployments in terms of relative importance of ISP and IXP footprint.
We start our sample of local ecosystems with the largest market of Netflix, the USA (Section~\ref{sec:deployment-us}). We follow with an emerging, though already large, market for Netflix, Brazil (Section~\ref{sec:deployment-br}). We finish the analysis with a look at Europe (Section~\ref{sec:deployment-eu}), exploring mature IXP ecosystems, as well as glimpse into the server deployment over time across different countries.
\begin{figure}[htb!]
	\includegraphics[width=\columnwidth]{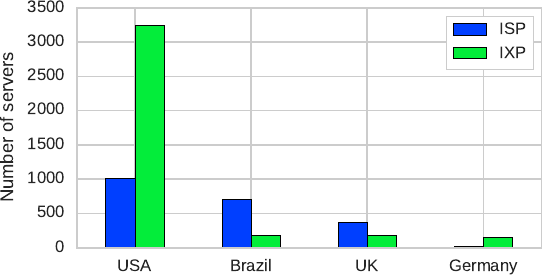}
	\caption{Deployment details for selected regions.}
	\label{fig:details-deployment}
\end{figure}
\subsection{USA}\label{sec:deployment-us}
We begin our look at local ecosystems with the United States of America. USA is the region with most Netflix customers by far~\cite{statista_netflix_subscribers2014}, and is supported by the largest server deployment of any country. Netflix has 3,236 IXP and 1,007 ISP servers deployed in the USA. Those servers are spread across 24 IXP and 205 ISP locations, reaching into 211 different ISPs.
We look first at the IXP deployment, given its numeric dominance (3,236 IXP vs. 1,007 ISP servers) for delivering content in the USA. Such a strong IXP deployment is perhaps surprising, given that according to the public information from PeeringDB, we find no American IXP in the top five of largest IXPs world-wide in terms of either members or capacity. Nonetheless, there is a significant number of IXPs across the country. Netflix has taken advantage of this footprint, and is present at 24 IXP locations (as identified by airport codes). The deployment covers the major metropolitan areas, picking the largest US IXPs according to PeeringDB member count. Netflix is present at nine of the ten largest IXPs in the USA, and 15 of the largest 20. 
Netflix's deployment at IXPs typically involves a significant number of servers, whereas deployment inside ISPs is more fine-grained. We encounter IXP deployments at 24 different locations, the largest consisting of 360 servers. For ISPs, the largest deployment in a single location consists of a mere 14 servers. However, ISP servers are installed at 205 locations in total. Deployment at ISPs therefore appears to complement the geographical reach of the IXP deployment, over a higher number of locations, but with relatively small deployment sizes at each location compared to IXP ones.
Note the absence\footnote{To discard the possibility of a measurement error, we included all reasonable abbreviations of these ISP names as input for the DNS crawler. However, even after this extensive search, we could not discover further servers.} of Netflix deployment inside four major ISPs (AT\&T, Comcast, Time Warner Cable and Verizon\footnote{We discovered three Netflix servers in Verizon’s network, which do not offer a significant advantage in traffic savings for such a large network, but might be part of a trial.}), as shown in Table~\ref{tab:isp-speedindex-us}. The explanation for this absence is that these ISPs publicly refused to deploy Netflix servers. Instead, they insisted on signing paid peering contracts with Netflix~\cite{netflix_net_neutrality, time_netflix_verizon}. This makes sense given the strong position of these ISPs in the US market.
ISP servers in the USA are hosted by smaller players. When contrasting ISPs with the Netflix ISP Speed Index\footnote{\url{http://ispspeedindex.netflix.com/}}, we observe that those ISPs which do not deploy servers provide similar performance results as those which have Netflix servers deployed. This suggests that deploying Netflix servers inside an ISP network does not automatically imply better performance, at least according to Netflix's own ISP Speed Index.
In summary, we observe that the USA has an IXP ecosystem mature enough, so that the available IXPs are sufficient for Netflix to rely primarily on IXPs to reach its large customer base. This comes in as a surprise, given that based on research literature little is known about the US IXP ecosystem, especially in comparison to European one ~\cite{Chatzis2015quovadis}. Furthermore, relying on IXP deployments, and not having deployments inside some ISPs, does not appear to have negative consequences on performance as reported by Netflix, highlighting again the usability of IXPs for large-scale content delivery.
\begin{table}
	\centering
	\begin{tabular}{lr|lr}
		\toprule
		\multicolumn{4}{c}{USA} \\
		\midrule
        AT\&T & - & Mediacom & - \\
        Bright House & - & Optimum & - \\
        CenturyLink & 113 / 11 & Spectrum & - \\
        Charter & - & Suddenlink & 68 / 31 \\
        Comcast & - & TWC & - \\
        Cox & - & Verizon & 3 / 2 \\
        Frontier & 19 / 3& Windstream & 31 / 11 \\
		\bottomrule
	\end{tabular}
	\caption{Netflix servers deployed inside US ISPs. ISPs are taken from Netflix's ISP Speed Index. The left number denotes the number of servers in an ISP, the right one the number of locations those servers are deployed at. ISPs listed multiple times in the index (e.g., due to different broadband connection types), are listed only once in this table.}
	\label{tab:isp-speedindex-us}
\end{table}
\subsection{Brazil}\label{sec:deployment-br}
Our second chosen local ecosystem is Brazil. This emerging market has the second largest Netflix server deployment despite not being an English-speaking country. Netflix offers in Brazil a substantial selection of content with at least subtitles in Portuguese for a fraction of the cost of cable TV. The deployment consists of 901 servers, 713 servers inside ISPs and 188 servers at IXPs. Unlike the USA, servers in Brazil are primarily located inside ISPs. ISP servers are deployed inside 187 ISPs, covering 58 locations across the vast Brazilian geography, but mostly along the Eastern coastal regions where most people live.
In strong contrast to the USA, IXP servers are only deployed at 3 locations on the South East Coast (São Paulo, Rio de Janeiro, and Porto Alegre) and at one location on the North East Coast (Fortaleza). In Brazil, Netflix has a limited IXP server deployment (in both locations and number of servers), despite a reasonably large number of available IXP locations (25 locations in total according to~\cite{pam2016br}, in the USA Netflix uses 24 IXP locations (Section~\ref{sec:deployment-us})). Deploying servers in IXPs has to be more cost efficient for Netflix due to economies of scale and a simpler contractual situation with fewer parties involved compared to ISP deployment. The observed deployment suggests an IXP ecosystem with limited capacity to reach Netflix customers. This limitation might be caused by a multitude of factors, including lack of capacity on the IXP switching fabric or the inability to host additional servers at IXPs.
According to PeeringDB data, the three IXPs on the South East Coast Netflix is present at, are also the largest ones, in terms of number of members. The IXP in Fortaleza is the seventh largest in Brazil. The Brazilian IXP infrastructure is developed by IX.br, a non-profit initiative. IX.br explicitly aims to improve the Internet connectivity deficiencies of the north, west and central regions, by providing a collection of exchange points. However, we see that Netflix only uses the IXP facilities at 3 (São Paulo, Rio de Janeiro, Fortaleza) of the 5 largest metropolitan areas, all located on the East coast. The vast majority of IXPs in Brazil have a small number of peers, and more importantly lack content providers, and private companies except in the South East~\cite{pam2016br}.
Brazil has a developing Internet infrastructure. External metrics such as the Netflix Speed Index figures show much lower bandwidth figures compared to the other top Netflix markets. Whereas IXPs by nature aim at fostering local access ecosystems, the edge Internet infrastructure must be strong enough for service providers to operate purely from these exchange points. Otherwise, deployment inside ISPs seems necessary.
\subsection{Europe}\label{sec:deployment-eu}
\begin{figure}
	\centering
	\includegraphics[width=\columnwidth]{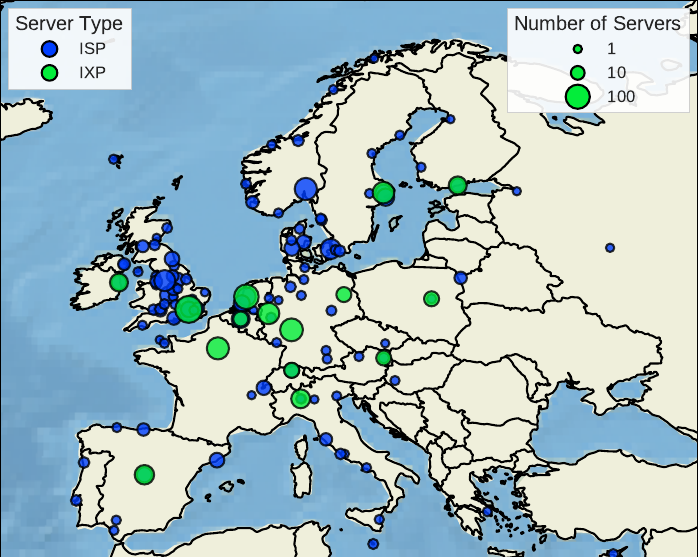}
	\caption{Netflix server deployment in Europe. Each marker denotes a location, the marker sizes are scaled by deployment size.}
	\label{fig:}
\end{figure}
\begin{figure}
	\centering
	\includegraphics[width=\columnwidth]{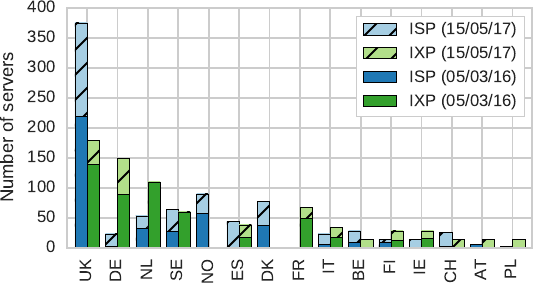}
	\caption{Netflix deployment inside European countries from May 3 2016 to May 15 2017. The upper, hatched parts of each bar indicate additional servers installed at the later time. Countries with more than 10 new servers represented.}
	\label{fig:netflix-time-dynamics}
\end{figure}

We continue our look at local ecosystems with the European region. Europe is a fragmented market, as most countries are rather small in terms of their population and geographic size, in comparison to countries such as the USA or Brazil. Europe is also different from other regions, in several respects. First, Europe has a diversity of languages, English being the primary language only in the UK and Ireland, and English literacy varies widely across European countries. Second, the ISP market in most European countries is still highly dominated by one incumbent player that owns the fiber infrastructure and is forced by law to give access to it to other ISPs. Finally, the European IXP market is very developed, having some of the largest IXPs in the world in terms of number of members. Indeed, four of the five largest IXPs in the world as measured by their number of members are European.
Netflix has adopted a country-specific approach for its European business expansion. In early 2012, it began to operate in UK and Ireland. Progressively, Netflix expanded to additional countries. When looking at individual countries, we expect to see a strongly IXP dominated deployment, given the maturity of European IXPs, as well as the limited ISP competition in many countries. In the remainder of this section, we explore the IXP presence in several countries with significant Netflix deployment, as well as the evolution of server deployment across Europe in order to obtain further insight in the European characteristics. 
\begin{table}
	\centering
	\begin{tabular}{lr|lr}
		\toprule
		\multicolumn{4}{c}{UK} \\
		\midrule
		BT & 107 / 6 & Sky & 82 / 20 \\ 
		EE & - & TalkTalk & 129 / 32 \\
		Plusnet & 6 / 1 & Virgin & 59 / 9 \\ 
		\bottomrule
	\end{tabular}
	\caption{Netflix servers deployed inside UK ISPs. The table follows the structure of Table~\ref{tab:isp-speedindex-us}.}
	\label{tab:isp-speedindex-uk}
\end{table}
\paragraph{The United Kingdom} The UK has the largest Netflix server deployment. We observe 554 servers in total, consisting of 375 servers deployed inside ISPs and 179 servers deployed at IXPs. All the IXP servers are deployed in the London area, where one of the largest IXPs in the world in terms of number of members, LINX, is present. Despite the size of LINX, and its ability to reach a large user base in London, Netflix unexpectedly relies on significant ISP deployment in the UK. ISP servers are deployed within 12 different networks at 35 different locations (see Table~\ref{tab:isp-speedindex-uk}). The predominance of ISP servers is likely due to the limited footprint of LINX across the UK outside London. Therefore, we argue the UK deployment illustrates (similar to the situation in Brazil) two points: (1) the presence of a large IXP is useful as much as it can reach Netflix's user base, but (2) without adequate footprint to cater the demand, the presence of a large IXP is insufficient and needs to be complemented by significant ISP deployment.
\begin{table}
	\centering
	\begin{tabular}{lr|lr}
		\toprule
		\multicolumn{4}{c}{Germany} \\
		\midrule
        EWE & 5 / 2 & Telekom & - \\
        Kabel Deutschland & - & Unitymedia & - \\
        M-net & 4 / 2 & Versatel & - \\
        NetCologne & 3 / 1 & Vodafone Germany & - \\
        o2 & - & ~ & ~ \\
		\bottomrule
	\end{tabular}
	\caption{Netflix servers deployed inside German ISPs. The table follows the structure of Table~\ref{tab:isp-speedindex-us}.}
	\label{tab:isp-speedindex-de}
\end{table}
\paragraph{Germany} Germany has a strong IXP ecosystem. Indeed, Germany hosts one of the largest IXPs in the world in terms of number of members: DE-CIX. However, contrary to the UK, Germany's native language is not English, despite a generally good English proficiency of its population. Netflix began operations in Germany in late 2014, and the current German market for Netflix is not that large. Netflix's German footprint consists of 149 IXP servers and only 23 ISP servers. Those ISP servers are deployed within 8 networks at 10 different locations, whereas the IXP servers are deployed in Frankfurt, Duesseldorf and Berlin. The predominantly IXP-based Netflix deployment in Germany indicates that, for now, the footprint of its IXPs is sufficient to fulfil the demand without much ISP deployment. As Germany still is a young market for Netflix, the future will have to show, to what extent the available IXPs will be used and when or where ISP deployment might become necessary. Further, as in the case of the USA, the lack of Netflix server deployment in some German ISPs does not seem to come with poor performance, as per the Netflix Speed Index. As in the USA, we notice the lack of deployment inside Deutsche Telekom, a tier-1 ISP and the incumbent operator in Germany. Consistently, as per Caida's AS relationship inference, Netflix is a customer of Deutsche Telekom.
\paragraph{Evolving Europe}
So far, our deployment study has focused on the current footprint of the Netflix infrastructure. However, Netflix deployment changes over time. Europe in particular has seen a staged expansion of Netflix to different countries, spanning from 2012 to 2016. In this section, we look at how server deployment across different European countries has changed. We will compare two snapshots, obtained on May 3 2016 and May 15 2017. Figure~\ref{fig:netflix-time-dynamics} shows per country server deployment at each snapshot. Solid bars indicate the footprint in 2016, whereas the overlapped hatched bars depict our 2017 measurements.
Comparing the two snapshots, we see that Netflix's footprint has grown significantly across Europe. We see multiple expansion patterns across the region; in some countries we observe that more IXP servers were added, whereas for other countries more ISP servers were added. In the later snapshot, we see that Netflix is present at IXPs in almost every European country, except Norway and Denmark.
The fragmentation of Europe compared to other regions has brought a relatively dense set of available IXP locations. These locations are convenient for a content provider to achieve significant reach across the region. We see in several large European countries (Germany, France, Italy and Spain) an initial IXP-only deployment, followed by ISP servers in the updated snapshot.
When looking at the six European countries where Netflix was first made available, five of them (GB, IE, SE, NO, NL) have expanded primarily by adding servers inside ISPs in our latest snapshot. We find such a development consistent with our findings so far: Whereas IXPs provide an effective platform to initially delivering content, ISP servers allow to achieve finer-grained penetration to reach the customer base. Please note, that such fine-grained ISP deployment, does not question the importance of IXPs for content delivery. In all such countries, where we observe additional ISP deployment, this happens in addition to and after the IXP deployment, not as a replacement for the IXP deployment.
When looking at the most recent markets for Netflix (Spain and Italy in 2015, Poland in 2016), we see as expected substantial development between these snapshots, taking into account how recently Netflix became present there. For these three countries, the current footprint is primarily supported by IXP servers, rather than ISP servers. The deployment in those countries reinforces the idea of a country expansion beginning at IXPs, followed later on with ISP deployments. France is another example of a substantial initial IXP deployment, with minimal ISP deployment (in this case a single server).
The changes over time in the deployment suggest a strategy where IXP locations are used to first establish a stronghold, possibly to test the demand in the region. Later on, more servers, also inside ISPs, are deployed, complementing the deployment in a fine-grained manner. These changes strengthen our findings on the strategic nature and importance of IXPs, for a player such as Netflix.
\subsection{Summary}
In this section, we have illustrated the diversity in local Internet ecosystems, as seen through the Netflix's server deployment. Our choice of local ecosystems has shown how the specifics of each local ecosystem translate into very different outcomes in terms of server deployment. We observed that ecosystems where developed IXPs are available typically lead to significant IXP server deployment. However, we also observed that to reach a large customer base, which is geographically scattered, ISP deployment is often necessary to compensate for the limited footprint of the local IXPs. When looking at the deployment over time, we observed multiple instances of Netflix first relying on IXP deployment, followed by ISP deployment. This suggests the use of IXPs as initial stronghold (when possible), and ISP deployment to improve the footprint.
We noticed quite a few instances of tier-1 ISPs, both in the USA and Europe, not deploying any Netflix servers. These are too many cases to be considered as pure coincidence. Indeed, to protect their tier-1 status, these ISPs may choose to treat Netflix as a customer and enforce paid peering, instead of deploying its servers. Such a strategy makes complete sense given that, apparently, it does not even come with a penalty in performance, as declared by the Netflix Speed Index.
 \section{Traffic}
\label{sec:traffic}
In this section, we complement the deployment footprint with estimates of the traffic sent by Netflix's servers. We first describe our methodology for estimating the traffic generated by each server (Section~\ref{sec:traffic_data_collection}). We then discuss our results, and compare them to the observed server deployment, thereby validating the footprint findings (Section~\ref{sec:traffic_traffic}).
\subsection{Data Collection Methodology}
\label{sec:traffic_data_collection}
The identification field (ID) in the IPv4 header is used to reassemble fragmented IP packets at the destination.
To do so, the ID field must be unique for each flow, within certain limitations.
Most operating systems use a global ID value, incremented for every packet sent~\cite{handley2001network}, irrespectively of the various flows kept by the end-host. The values of the ID field therefore provide a proxy measurement for the traffic volume generated by a device. However, this only holds true if the operating system consistently increments the ID value for each packet sent. The specific numbering restriction is not guaranteed as per RFC6864~\cite{rfc6864}, which also allows alternative schemes, such as using arbitrary or random values.
Fortunately, at the time we performed this study, all Netflix servers ran FreeBSD 10~\cite{netflix_oc_software}.
FreeBSD 10.3, the production release at this time, with default settings, generates IDs in a predictable way.
In particular, the ID values for both ICMP and TCP packets are derived from the same counter. Accordingly, we can estimate the total amount of traffic sent from a server by issuing ICMP ping requests and evaluating the ID values in the replies.
We ran a measurement campaign to assess the temporal behaviour of the ID field of Netflix's servers.
Due to the limited size of the ID field (16 bit), overflows of the ID field will happen, especially on busier servers.
To be able to detect and correct such overruns, we sent ping requests every 30ms.
Because the network load induced by simultaneously sampling all servers would be prohibitive, we sampled each server for one minute in a round-robin fashion, using 150 hping3~\cite{hping} processes.
This led to one measurement per server approximately every 30 minutes.
We ran the measurements for a period of 10 days, starting November 21 2016.
The traffic load generated by targeting each server twice in an hour is small enough to be negligible for Netflix, and hence not disturbing its regular business activities.
From our measurements, we estimate how much traffic is served by the Netflix infrastructure.
We assume that most packets use the standard Ethernet Maximum Transmission Unit (MTU) of 1500 bytes:
To make better use of the available link capacity, servers tend to make individual packets as large as possible.
This should particularly be the case for video traffic, where the content is large in size.
Most of the IXP peering LANs Netflix is connected to also use an MTU of 1500 bytes, further supporting our choice.
We therefore estimate the total traffic by multiplying the observed packet rates with this MTU size.\footnote{While this is an assumption, we are confident that at least the distribution of MTU sizes is similar across regions and servers, so that the trends are correct, even if the exact numbers are not.}\footnote{Due to packet fragmenation (for example in cellular networks), the individual fragments sent might be smaller than this MTU. However, all fragments stemming from the same packet will carry the same IP ID, making this apporach resistent to packet fragmentation.}\footnote{Servers might also send out types of messages not directly related to video delivery (e.g. for management purposes), leading to an overestimate of the video traffic sent. However, as these servers are explicitly built to deliver video content, the vast majority of the traffic should be video content, making the impact of other traffic types negligable.} 
\subsection{Traffic follows deployment}
\label{sec:traffic_traffic}
\begin{figure}
    \includegraphics[width=\columnwidth]{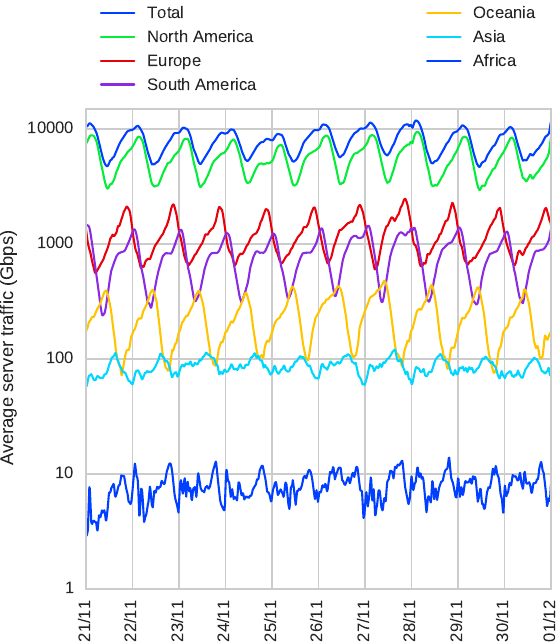}
    \caption{Average traffic per continent over all IXP and ISP servers. Continents in the legend are ordered by decreasing volume of traffic. Time is depicted in UTC.}
    \label{fig:traffic-per-continent-time}
\end{figure}

\begin{figure}
    \includegraphics[width=\columnwidth]{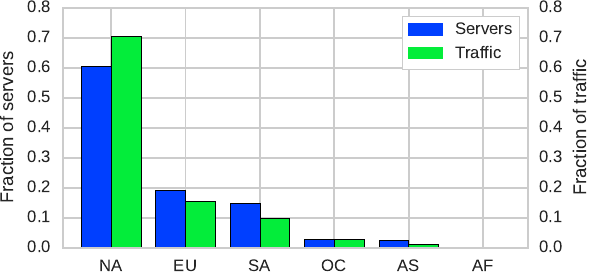}
    \caption{Distribution of Netflix servers and traffic per continent.}
    \label{fig:servers-per-continent}
\end{figure}

\begin{figure}
    \includegraphics[width=\columnwidth]{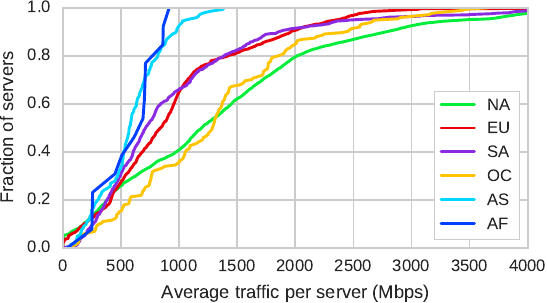}
    \caption{Average traffic generated per server for each continent.}
    \label{fig:traffic-per-server-continent}
\end{figure}

\begin{figure}
    \includegraphics[width=\columnwidth]{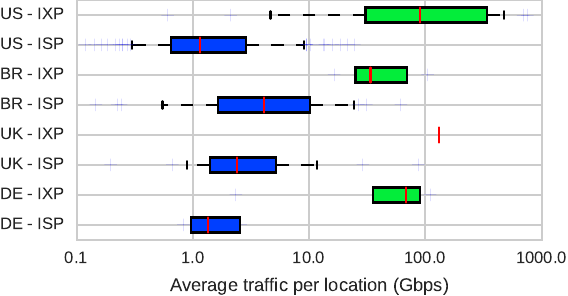}
    \caption{Traffic generated at ISP and IXP locations per country. Locations are inferred from airport codes.}
    \label{fig:traffic-boxplot}
\end{figure}

\begin{figure}
    \includegraphics[width=\columnwidth]{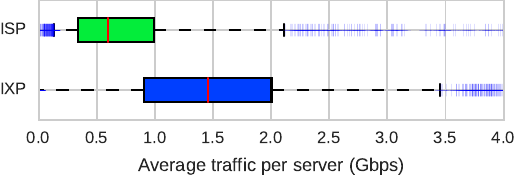}
    \caption{Average traffic generated by ISP and IXP servers globally.}
    \label{fig:traffic-per-type-global}
\end{figure}

Figure~\ref{fig:traffic-per-continent-time} shows a time series of the average Netflix traffic per continent, measured in Gbps. The plot clearly exposes a diurnal traffic pattern. A closer investigation reveals that the daily peaks per continent are shifted consistently with their geographic location. We can also see that during Thanksgiving Day (24th of November)\footnote{Note that the plot itself is in UTC, hence the trove appears to be at November 25.}, traffic in North America experienced a significant decrease of traffic during peak time. These observations match our expectations for a large-scale video delivery service, making us confident that the measurement technique provides us with reliable traffic figures.\footnote{Up to a certain scaling factor, induced by our choice of packet sizes, as described in Section~\ref{sec:traffic_data_collection}.}
When looking at the amount of traffic in Figure~\ref{fig:traffic-per-continent-time}, we see that global Netflix traffic peaks at 11.8 Tbps. North America is the dominating continent in terms of traffic. South America and Europe are comparable, followed by Oceania and Asia. In Africa we observe very little Netflix traffic.
The continent ranking derived from our traffic observations is consistent with the one from the server deployment sizes.
Figure~\ref{fig:servers-per-continent} shows a bar plot comparing the relative deployment sizes per continent with the relative amount of traffic generated.
The continent with the largest deployment, North America, also generates the most traffic.
Europe and South America generate comparable amounts of traffic, consistently with their similar deployment sizes.
Asia and Australia exhibit significantly smaller deployments and also smaller traffic volumes.
We now zoom into the per-server traffic. Figure~\ref{fig:traffic-per-server-continent} shows the distribution of per-server traffic for each continent. There is significant variety in the traffic generated by individual servers. Servers deployed in North America, Europe, South America and Oceania have a similar range of traffic values. Nevertheless, servers in North America and Oceania have higher traffic per server, when compared to South America and Europe. In contrast, servers deployed in Asia and Africa exhibit noticeably lower average and maximum traffic values. In those regions with bigger customer bases, i.e. the Americas, Europe and Oceania, we observe a higher server utilization than in Asia and Africa, where the customer base is smaller.
We finally contrast our previous observations about regional ecosystems with our traffic measurements. Figure~\ref{fig:traffic-boxplot} shows a boxplot summarizing traffic measured from IXP and ISP locations in the USA, Brazil, UK and Germany. We confirm in all four countries that IXP locations deliver substantially more traffic than the ISP locations. We had observed a variety of IXP to ISP server deployment ratios in these countries. The observations from the traffic further support our previous observations. These traffic figures shows that IXP locations act as large hubs, which are complemented by a fine-grained deployment of ISP servers. Whereas every country shows different traffic figures, we observe about an order of magnitude more traffic from IXP locations.
This also holds true when looking at individual servers grouped by type. Figure~\ref{fig:traffic-per-type-global} shows that the median IXP server is responsible for roughly three times as much traffic as the median ISP server. Further, when considering the distribution of traffic values between the lower and upper quartiles, as depicted by the green and blue boxes in the plot, the relationship remains the same. This finding underpins the importance of IXP server deployment for the delivery infrastructure, as not just only the majority of servers is deployed at IXPs, but they also deliver more traffic then their ISP counterparts, enabling Netflix to deliver the majority of its traffic from IXP locations. The combined worldwide IXP substrate thus is the major corner stone of the Open Connect CDN.
\subsection{Summary}
In summary, we observe that the server deployment sizes and the corresponding traffic estimates are consistent. The traffic per continent is consistent with our expectations of the relative market sizes. Further, we observed per-server average traffic volumes in the Gbps ranges, with heavier loads for IXP servers compared to ISP ones, again highlighting the importance of IXPs for a global content player like Netflix.
 \section{Related Work}
\label{sec:related}
As one of the major players in video content delivery, Netflix's role in the Internet directly illustrates the observations from Labovitz et al.~\cite{labovitz2010internet}, back in 2010. 
Indeed, Labovitz et al.~\cite{labovitz2010internet} observed a new trend, whereby traffic was seen to flow directly between large content providers, datacenters, CDNs and consumer networks, away from large transit providers. 
Subsequent studies investigated the potential implications of more direct interconnections on the Internet~\cite{castro2014remote,dhamdhere2010internet,dhamdhere2011twelve,ma2015evolution}.
Due to the success of players such as Netflix, the rise in video traffic observed by Labovitz et al.~\cite{labovitz2010internet} has only continued. Our study of the server deployment of Netflix at the edge of the Internet, and the corresponding traffic delivered to end-users, makes the observations of Labovitz et al.~\cite{labovitz2010internet} even more relevant today.
Despite their importance in the Internet ecosystem, only a few studies have targeted IXPs~\cite{Ager2012IXP,augustin2009ixps,Cardona2012IXP,Chatzis2013benefits,Chatzis2013meets,Chatzis2015quovadis} and their role in the Internet. The work from Augustin et al.~\cite{augustin2009ixps} aimed at systematically mapping IXP infrastructures through large-scale active measurements, leading to the first evidence of the huge number of IXPs around the world. Ager et al.~\cite{Ager2012IXP} studied the ecosystem and traffic of one of the largest European IXPs, while Restrepo et al.~\cite{Cardona2012IXP} looked at two smaller European IXPs. Subsequent studies from Chatzis et al.~\cite{Chatzis2013benefits,Chatzis2013meets,Chatzis2015quovadis} reinforced the critical role played by IXPs in the Internet ecosystem.
IXPs are a major component supporting the peering ecosystem of the Internet. To this day, however, the role of IXPs world-wide in supporting the delivery of large amounts of traffic close to end-users has been understated. Indeed, despite the large number of IXPs known to exist~\cite{augustin2009ixps}, the largest of them having hundreds of members and delivering daily traffic volumes in the petabyte range, their relative importance for content delivery was largely unreported. In this work, we uncovered and quantified the importance that IXPs play in enabling a player such as Netflix to deliver its traffic to its large and worldwide customer base. We observed that despite preferring to deploy servers within ISP networks, a majority of Netflix servers and the corresponding traffic exploit the strategic location and ecosystems provided by IXPs all around the world. Labovitz et al.~\cite{labovitz2010internet} indicated a significant shift in the mental map of the Internet, with traffic being increasingly delivered directly between large content providers and consumer networks, away from large transit providers. Our work adds another piece of evidence for this shift, with a direct observation of a large video delivery provider doing this by strategically exploiting the rich ecosystem that many IXPs provide. 
Mapping the server deployment and expansion of a large content player has been done before. Calder et al.~\cite{calder2013mapping} developed techniques that enumerate IP addresses of servers of the Google  infrastructure, found their geographic location, and identified the association between clients and clusters of servers. To do this accurately, they use the EDNS-client-subnet DNS extension to measure which clients a service maps to which of its serving sites. Different from our work, Calder et al.~\cite{calder2013mapping} focused on the accuracy of the server mapping and geolocation, necessary given the size and complexity of the Google infrastructure. In this paper, we focus on the types of locations where Netflix has chosen to deploy its server infrastructure. Further, different from Calder et al.~\cite{calder2013mapping}, we provide estimates of the traffic delivered by the Netflix servers. Overall, we are not overly concerned with the mapping of the servers itself, as Netflix runs a single service, contrary to Google. Rather, our focus is on the implications of Netflix's server deployment strategy, with the lens it provides on the Internet ecosystem.
 \section{Discussion}
\label{sec:discussion}
In this section we will discuss our most important findings regarding the current state of the IXP ecosystem and its usability as a base for content delivery.
One peculiarity of the way Netflix delivers its content, is that, in contrast to the other big video players by traffic volume (YouTube and Amazon Video), it does so without operating a backbone network~\cite{netflix_peering_locations}. To reach its customers, Netflix instead relies on deploying servers at IXPs and inside ISPs. These deployment sites form self-sufficient islands, capable of serving the local customer demand more or less independently. Netflix's pre-fetching approach to populate content on its servers is key to reduce the amount of transit traffic, i.e., traffic between the servers holding the original content and the copies placed on the deployment sites. The backbone-less and light in transit approach of Netflix contributes to the observed phenomenon of Internet flattening. Instead of flowing through the traditional Internet hierarchy (tier-1s), Internet traffic goes through more and larger direct interconnects between networks at the edge. To deliver its traffic, Netflix chooses IXP locations, as well as ISPs that are not in the traditional core of the Internet, therefore bypassing the traditional Internet hierarchy and inevitably contributing to the observed flattening.
The case of Netflix demonstrates that large-scale traffic delivery from edge locations (esp. IXP locations) is possible. We believe that reporting this approach followed by Netflix is important, as it illustrates its feasibility, but also the challenges that come with it, in terms of being able to exploit the very different local ecosystems of the Internet. This will hopefully inspire other small and large players to follow a similar approach, at least for some parts of their content, which then may in turn exacerbate the flattening phenomenon.
Netflix not only does not operate a backbone, but it nowadays also does not operate a single datacenter either~\cite{netflix_cloud_migration}. Instead, Netflix serves its traffic from servers deployed in colocation housing locations at or in close proximity to IXPs. These locations allow Netflix to operate without its own datacenters, as those locations essentially provide all the features of a regular datacenter. One drawback of such an approach is the space restrictions in these locations that might limit their usability for large deployments. Nevertheless, for Netflix's needs focused on data storage and data transfer, not operating its own datacenters seems to work.
To our knowledge, it is the first time such a worldwide deployment is exposed, based on a strategic use of IXP facilities as a datacenter replacement. From this, we learn that the benefit of IXPs is not limited to network interconnection~\cite{Ager2012IXP}, but that they also facilitate the deployment of large server bases at locations with strategically beneficial network connectivity.
 \section{Summary}
\label{sec:summary}
In this work, we studied the global footprint of one content hypergiant, Netflix, to gain a new perspective on the current Internet. We exposed the approach used by Netflix to deliver massive amounts of traffic from over 500 world-wide locations with neither a backbone nor datacenters. It does so by deploying its own servers at IXP locations as well as in ISP networks. By studying the deployment of its servers, we highlighted regional differences in the deployment, by sampling the diversity of local ecosystems that collectively make up the Internet. The Netflix lens provides evidence for the vastly understated ability of the many IXPs world-wide to deliver large amounts of traffic on a global scale. The world-wide footprint of IXPs is the major corner stone of Open Connect and enables Netflix to operate a global content delivery system, with very limited transit traffic, and without operating a backbone or owning datacenters.
\bibliographystyle{abbrv}
\bibliography{netflix}

\begin{thebibliography}{10}

\bibitem{adhikari2015measurement}
V.~K. Adhikari, Y.~Guo, F.~Hao, V.~Hilt, Z.-L. Zhang, M.~Varvello, and
  M.~Steiner.
\newblock {Measurement Study of Netflix, Hulu, and a Tale of Three CDNs}.
\newblock {\em IEEE/ACM ToN}, 23(6):1984--1997, 2015.

\bibitem{Ager2012IXP}
B.~Ager, N.~Chatzis, A.~Feldmann, N.~Sarrar, S.~Uhlig, and W.~Willinger.
\newblock {Anatomy of a Large European IXP}.
\newblock In {\em Proc. of ACM SIGCOMM}, 2012.

\bibitem{augustin2009ixps}
B.~Augustin, B.~Krishnamurthy, and W.~Willinger.
\newblock {IXPs: Mapped?}
\newblock In {\em Proc. of ACM IMC}, 2009.

\bibitem{google_ct_symantec}
G.~S. Blog.
\newblock {Sustaining Digital Certificate Security}.
\newblock
  \url{https://security.googleblog.com/2015/10/sustaining-digital-certificate-security.html},
  2015.

\bibitem{draft-ietf-dnsop-nxdomain-cut}
S.~Bortzmeyer and S.~Huque.
\newblock {NXDOMAIN: There Really Is Nothing Underneath}.
\newblock RFC 8020, IETF, 2016.
\newblock \url{https://tools.ietf.org/html/rfc8020}.

\bibitem{pam2016br}
S.~H.~B. Brito, M.~A. Santos, R.~dos Reis~Fontes, D.~A.~L. Perez, and C.~E.
  Rothenberg.
\newblock {Dissecting the Largest National Ecosystem of Public Internet
  eXchange Points in Brazil}.
\newblock In {\em Proc. of PAM}, 2016.

\bibitem{caida_as_rank}
CAIDA.
\newblock {AS Rank}.
\newblock \url{http://as-rank.caida.org/}.

\bibitem{calder2013mapping}
M.~Calder, X.~Fan, Z.~Hu, E.~Katz-Bassett, J.~Heidemann, and R.~Govindan.
\newblock {Mapping the Expansion of Google's Serving Infrastructure}.
\newblock In {\em Proc. of ACM IMC}, 2013.

\bibitem{Cardona2012IXP}
J.~C. Cardona~Restrepo and R.~Stanojevic.
\newblock {IXP Traffic: A Macroscopic View}.
\newblock In {\em Proc. of ACM LANC}, 2012.

\bibitem{castro2014remote}
I.~Castro, J.~C. Cardona, S.~Gorinsky, and P.~Francois.
\newblock {Remote Peering: More Peering Without Internet Flattening}.
\newblock In {\em Proc. of ACM CoNEXT}, 2014.

\bibitem{certificate_transparency}
{Certificate Transparency}.
\newblock \url{https://www.certificate-transparency.org}.

\bibitem{Chatzis2013benefits}
N.~Chatzis, G.~Smaragdakis, J.~B\"{o}ttger, T.~Krenc, and A.~Feldmann.
\newblock {On the Benefits of Using a Large IXP as an Internet Vantage Point}.
\newblock In {\em Proc. of ACM IMC}, 2013.

\bibitem{Chatzis2013meets}
N.~Chatzis, G.~Smaragdakis, A.~Feldmann, and W.~Willinger.
\newblock {There is More to IXPs Than Meets the Eye}.
\newblock {\em ACM CCR}, 43(5), Nov. 2013.

\bibitem{Chatzis2015quovadis}
N.~Chatzis, G.~Smaragdakis, A.~Feldmann, and W.~Willinger.
\newblock {Quo Vadis Open-IX?}
\newblock {\em ACM CCR}, 45(1), Jan. 2015.

\bibitem{teamcymru_ipasnmapping}
T.~Cymru.
\newblock {IP to ASN Mapping}.
\newblock \url{http://www.team-cymru.org/IP-ASN-mapping.html}.

\bibitem{dhamdhere2010internet}
A.~Dhamdhere and C.~Dovrolis.
\newblock {The Internet Is Flat: Modeling the Transition from a Transit
  Hierarchy to a Peering Mesh}.
\newblock In {\em Proc. of ACM CoNEXT}, 2010.

\bibitem{dhamdhere2011twelve}
A.~Dhamdhere and C.~Dovrolis.
\newblock {Twelve Years in the Evolution of the Internet Ecosystem}.
\newblock {\em IEEE/ACM ToN}, 19(5):1420--1433, 2011.

\bibitem{statista_netflix_subscribers2014}
{Digital TV Research}.
\newblock {Number of Netflix paying streaming subscribers in 3rd quarter 2014,
  by country (in 1,000s)}.
\newblock
  \url{https://www.statista.com/statistics/324050/number-netflix-paying-streaming-subscribers/},
  2014.

\bibitem{gill2008flattening}
P.~Gill, M.~Arlitt, Z.~Li, and A.~Mahanti.
\newblock {The Flattening Internet Topology: Natural Evolution, Unsightly
  Barnacles or Contrived Collapse?}
\newblock In {\em Proc. of PAM}. 2008.

\bibitem{handley2001network}
M.~Handley, V.~Paxson, and C.~Kreibich.
\newblock {Network Intrusion Detection: Evasion, Traffic Normalization, and
  End-to-End Protocol Semantics.}
\newblock In {\em Proc. of USENIX Security Symposium}, 2001.

\bibitem{hping}
{Hping - Active Network Security Tool}.
\newblock \url{http://hping.org/}.

\bibitem{labovitz2010internet}
C.~Labovitz, S.~Iekel-Johnson, D.~McPherson, J.~Oberheide, and F.~Jahanian.
\newblock {Internet inter-domain traffic}.
\newblock {\em Proc. of ACM SIGCOMM}, 2010.

\bibitem{ma2015evolution}
R.~T. Ma, J.~Lui, and V.~Misra.
\newblock {Evolution of the Internet Economic Ecosystem}.
\newblock {\em IEEE/ACM ToN}, 23(1):85--98, 2015.

\bibitem{netflix_oc_software}
{Netflix}.
\newblock {Appliance Software | Open Connect}.
\newblock \url{https://openconnect.netflix.com/en/software/}.

\bibitem{netflix_cloud_migration}
{Netflix}.
\newblock {Completing the Netflix Cloud Migration}.
\newblock
  \url{https://media.netflix.com/en/company-blog/completing-the-netflix-cloud-migration/}.

\bibitem{netflix_oc_fill}
{Netflix}.
\newblock {Fill, Updates, and Maintenance | Open Connect}.
\newblock \url{https://openconnect.netflix.com/en/fill/}.

\bibitem{netflix_net_neutrality}
{Netflix}.
\newblock {Internet Tolls And The Case For Strong Net Neutrality}.
\newblock
  \url{https://media.netflix.com/en/company-blog/internet-tolls-and-the-case-for-strong-net-neutrality/}.

\bibitem{netflix_openconnect_portal_naming}
{Netflix}.
\newblock {Partner Portal Naming Conventions}.
\newblock \url{https://openconnect.netflix.com/en/portal-naming}.

\bibitem{netflix_peering_locations}
{Netflix}.
\newblock {Peering Locations | Open Connect}.
\newblock \url{https://openconnect.netflix.com/en/peering-locations/}.

\bibitem{netflix_openconnect}
{Netflix Open Connect}.
\newblock \url{https://openconnect.netflix.com/}.

\bibitem{netflix_openconnect_blog_map}
{How Netflix Works With ISPs Around the Globe to Deliver a Great Viewing
  Experience}.
\newblock
  \url{https://media.netflix.com/en/company-blog/how-netflix-works-with-isps-around-the-globe-to-deliver-a-great-viewing-experience}.

\bibitem{netflix_openconnect_deploymentguide}
{Netflix OpenConnect Appliance Deployment Guide}.
\newblock \url{http://oc.nflxvideo.net/docs/OpenConnect-Deployment-Guide.pdf}.

\bibitem{netflix_countries}
{Where is Netflix available?}
\newblock \url{https://help.netflix.com/en/node/14164}.

\bibitem{peeringdb}
{PeeringDB}.
\newblock \url{https://www.peeringdb.com}.

\bibitem{rfc2818}
E.~Rescorla.
\newblock {HTTP Over TLS}.
\newblock RFC 2818, IETF, 2000.
\newblock \url{https://tools.ietf.org/html/rfc2818}.

\bibitem{time_netflix_verizon}
{Verizon Won't Use Netflix's Hardware to Boost Streaming Speeds}.
\newblock \url{http://time.com/2866004/verizon-netflix/}.

\bibitem{rfc6864}
J.~Touch.
\newblock {Updated Specification of the IPv4 ID Field}.
\newblock RFC 6864, IETF, 2013.
\newblock \url{https://tools.ietf.org/html/rfc6864}.

\end{thebibliography}
\end{document}